\documentclass{ws-ijmpa}
\usepackage[super,compress]{cite}
\usepackage{graphicx}
\usepackage{hyperref}

\usepackage{xspace}
\usepackage {placeins}
\usepackage{tikz}
\usepackage{mathtools}
\DeclarePairedDelimiter{\abs}{\lvert}{\rvert}

\newcommand{\pythiaeight}{\textsc{Pythia}\,8\xspace}
\newcommand{\mg}{Madgraph\xspace}
\def\pT{\ensuremath{p_\mathrm{T}}\xspace}
\def\kt{\ensuremath{k_t}\xspace}

\newcommand{\Nchg}{\ensuremath{N_\text{ch}}\xspace}
\newcommand{\ZpT}{\ensuremath{p_\mathrm{{T}}^\mathrm{{Z}}}\xspace}

\def\Z{$Z$-boson\xspace}

\newcommand{\FigRef}[1]{Figure~\ref{#1}\xspace}

\begin{document}

\thispagestyle{plain}

\def\bib{B\kern-.05em{I}\kern-.025em{B}\kern-.08em}
\def\btex{B\kern-.05em{I}\kern-.025em{B}\kern-.08em\TeX}

\markboth{Kar and Rafanoharanara}{Probing underlying event in Z-boson events using event shape observables}

%
\catchline{}{}{}{}{}
%

\title{Probing underlying event in Z-boson events using event shape observables
}

\author{Deepak Kar}
\address{School of Physics and Institute for Collider Particle Physics, 
University of Witwatersrand\\
Johannesburg, South Africa.\\
deepak.kar@cern.ch}

\author{Dimbiniaina Soanasolo Rafanoharanara~\footnote{Formerly AIMS, Cape Town, South Africa}}
\address{School of Physics, 
University of Witwatersrand\\
Johannesburg, South Africa\\
dimbi@aims.ac.za}

\maketitle

\begin{history}
\received{\today}
\end{history}

\begin{abstract}
Experimental measurements of observables sensitive to the underlying event (UE) in $Z$-boson events
have been performed by both ATLAS and CMS experiments at the LHC. 
However, in the busy LHC environment, these observables receive substantial 
contribution from jets originating from initial state radiation (ISR). 
We probe if using event shape observables in conjunction
with the UE observables can help us to disentangle the effect of the UE from jets originating in ISR.

\end{abstract}

\ccode{PACS numbers: 12.38.-t, 13.85.-t}


\section{Introduction}
\label{sec:intro}

The underlying event (UE) is defined as the activity accompanying the hard-scattering in a hadronic collision event.
This includes hadronised decay products from partons not participating in a hard-scattering 
process (beam remnants), 
and additional scattering events in the same proton-proton collision, 
termed multiple parton interactions (MPI).
Initial- and final-state gluon radiation (ISR, FSR) also 
contribute to the UE activity. 
The soft interactions contributing to the UE cannot be calculated reliably using
perturbative Quantum Chromodynamics (pQCD) methods, and
are generally described using different phenomenological models, usually implemented
in Monte Carlo (MC) event generators. 
It is impossible to unambiguously separate the UE from the
hard scattering process on an event-by-event basis. However, distributions have been
measured that are sensitive to the properties of the UE. 
These measurements are critical inputs
to the tuning of MC event generators, especially of the parameters controlling different aspects
of soft QCD.

\begin{figure}[tbp]
  \begin{center}
    \begin{tikzpicture}[>=stealth, very thick, scale=1.1]
      \small

      \draw[color=blue!80!black] (0, 0) circle (3.0);
      \draw[rotate= 30, color=gray] (-3.0, 0) -- (3.0, 0);
      \draw[rotate=-30, color=gray] (-3.0, 0) -- (3.0, 0);

      \draw[->, color=black, rotate=-2] (0, 3.5) arc (90:47:3.5) node[right] {$\Delta{\phi}$};
      \draw[->, color=black, rotate=2] (0, 3.5) arc (90:133:3.5) node[left] {$-\Delta{\phi}$};

      \draw[->, color=red, ultra thick] (0, 2) -- (0, 4) node[above] {\textcolor{black}{\Z}};
      \draw[->, color=green!70!black, ultra thick] (0, -2) -- ( 0.0, -4);
      \draw[->, color=green!70!black, ultra thick, rotate around={-15:(0,-0.3)}] (0, -2) -- ( 0.0, -4);
      \draw[->, color=green!70!black, ultra thick, rotate around={ 15:(0,-0.3)}] (0, -2) -- ( 0.0, -4);

      \draw (0,  1.4) node[text width=2cm] {\begin{center} Toward \goodbreak $|\Delta\phi| < 60^\circ$ \end{center}};
      \draw (0, -1.1) node[text width=2cm] {\begin{center} Away \goodbreak $|\Delta\phi| > 120^\circ$ \end{center}};
      \draw ( 1.7, 0.3) node[text width=3cm] {\begin{center} Transverse \goodbreak $60^\circ < |\Delta\phi| < 120^\circ$ \end{center}};
      \draw (-1.7, 0.3) node[text width=3cm] {\begin{center} Transverse \goodbreak $60^\circ < |\Delta\phi| < 120^\circ$ \end{center}};
    \end{tikzpicture}
    \caption{Definition of the UE regions as a function of the azimuthal angle with respect to the direction of the \Z.}
    \label{fig:ueregions}
  \end{center}
\end{figure}
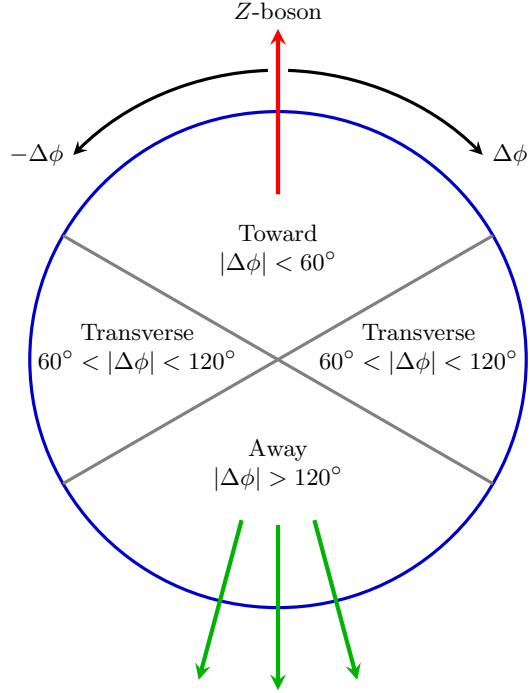

Measurements of such distributions have been performed at the Tevatron and at the LHC in several
different final states, including that in $Z$-boson events~\cite{Aaltonen:2010rm, Aad:2014jgf, Chatrchyan:2012tb, Sirunyan:2017vio}.
Since there is no final-state gluon radiation associated with a \Z , 
lepton-pair production consistent with \Z decays provides a cleaner final-state environment than jet production
for measuring the characteristics of the underlying event in certain regions of phase space.
The direction of the \Z candidate is used to define regions in the azimuthal plane
that have different sensitivity to the UE, a concept first used in Ref.~\cite{Field:2002vt}.
As illustrated in \FigRef{fig:ueregions}, the azimuthal angular difference~\footnote{
The standard collider reference system is a Cartesian right-handed coordinate system, with the nominal collision point at the origin. 
The anti-clockwise beam direction defines the positive $z$-axis, while the positive $x$-axis is defined as pointing from the collision 
point to the center of the ring and the positive $y$-axis points upwards. The azimuthal angle $\phi$ is measured
around the beam axis, and the polar angle $\theta$ is the angle measured with respect to~the $z$-axis. 
The pseudorapidity is given by $\eta = -\ln\tan\mspace{-0.1mu}( \theta/2 )$. 
Transverse momentum, \pT is defined relative to the beam axis.}
between charged tracks and the \Z,
$|\Delta\phi|=|\phi-\phi_\text{\Z}|$, is used to define the following three azimuthal UE regions:
\begin{itemize}
\item $|\Delta\phi| < 60^{\circ}$, the \textit{toward} region,
\item $60^{\circ} < |\Delta\phi| < 120^{\circ}$, the \textit{transverse} region, and
\item $|\Delta\phi| > 120^{\circ}$, the \textit{away} region.
\end{itemize}

The away region is dominated by particles balancing the momentum of the \Z except at low values of \ZpT .
The transverse region is considered as sensitive to the underlying event, since it is by construction
perpendicular to the direction of the \Z and hence it is expected to have a lower
level of activity from the hard scattering process compared to the away region.

In the busy LHC environment, the production of additional jets from hard radiation contaminates 
the transverse regions, as a result the transverse regions are no longer sensitive to only the UE.
In order to mitigate that, the two opposite transverse regions may be distinguished on an event-by-event basis through their amount of activity, 
as measured by the scalar sum of the charged-particle transverse momenta 
in each of them.
The more or less-active transverse regions are then referred to as \textit{trans-max}
and \textit{ trans-min}, respectively~\cite{PhysRevD.38.3419, PhysRevD.57.5787}. 
Then the trans-max region
is expected to contain the contribution from jets originating from ISR (will be referred to as ISR jets subsequently for brevity)
whereas the trans-min region is expected to be more sensitive to the UE.
The activity in the toward region is similarly expected to be
unaffected by additional activity from the hard
scatter.

\begin{figure}[h!]
  \centering
   \includegraphics[width=0.6\textwidth]{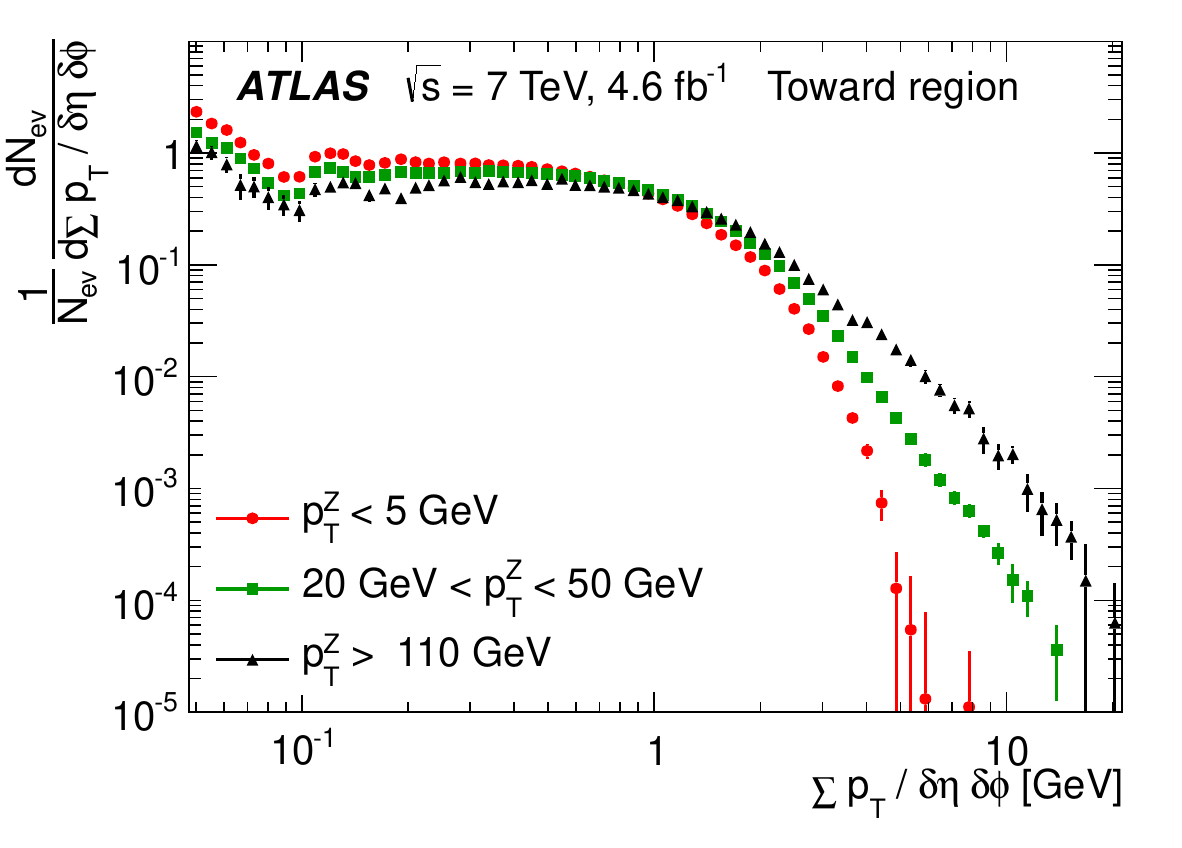} \\
  \caption{Distributions of the scalar \pT\ sum density of charged particles, for three different
  \Z transverse momentum (\ZpT) intervals, in the toward region~\protect\cite{Aad:2014jgf} 
  (ATLAS Experiment \copyright~2018 CERN).}
  \label{fig:atlas1d}
\end{figure}

However, even that approach has its limitations, as can be demonstrated by the distribution of \pT
of charged particles in the toward region in $Z$-boson events, shown in \FigRef{fig:atlas1d}.
This analysis selects $Z$-boson events decaying to two oppositely charged, same-flavour leptons.
If there were no ISR jets contributing
to the toward region, the activity would have been the same in all the \ZpT ranges, which is clearly not the case. A similar
result was obtained in the trans-min region as well. 

In this paper, we propose that by using event topology, we can identify a region of phase space for $Z$-boson events, where
the contamination from ISR jets is minimal, and which hence can be used to measure the UE activity more cleanly. Transverse event shape
observables~\cite{Berger:2010xi,Banfi:2010xy} are used to describe the event topology. 
The event shape observables in $Z$-boson events have been
measured as well at the LHC~\cite{Aad:2016ria, Chatrchyan:2013tna},
and they have been shown
to be sensitive to MPI~\cite{Cuautle:2014yda} in inclusive inelastic events.

\section{Definition of the observables}
\label{sec:obs}

The UE observables measured in this analysis are derived from the number, \Nchg , and transverse momenta, \pT ,
of stable charged particles in each event. These are defined for each azimuthal region under consideration.
They are normalised to per unit $\eta$--$\phi$ area, in order to be independent of the measurement $\eta$ range,
and to facilitate direct comparison between different azimuthal areas.

The event shape observables considered are the transverse thrust and the transverse sphericity.
The transverse thrust for a given event is defined as:
\begin{equation}
 \mathcal{T}_{\textrm{T}}= \max_{\hat{n}} \frac{\displaystyle\sum_{\textrm{i}} \abs{\vec{p}_{\textrm{T,i}}\cdot\hat{n}}}{\displaystyle\sum_{\textrm{i}} \abs{\vec{p}_{\textrm{T,i}}}}
\end{equation}
where the sum is performed over all the particles in an event and
$\vec{p}_{\textrm{T,i}}$ are the transverse momenta of the i$^{\textrm{th}}$ particle.
The unit vector $\hat{n}$ that maximizes the sum ratio is called the thrust
axis, $\hat{n}_T$.
Transverse thrust ranges from $\mathcal{T}_{\textrm{T}} = 1$  for a back-to-back and
to $\mathcal{T}_{\textrm{T}} = 2/\pi (\langle \abs{\cos\theta} \rangle)$ for a circularly symmetric
 distribution of particles in the transverse plane, respectively.

Sphericity is defined to describe isotropy of energy flow. It is based on the
quadratic momentum tensor:
\begin{equation}
	S^{\alpha\beta} = \frac{\sum_{\textrm{i}} p^{\alpha}_{\textrm{i}} p^{\beta}_{\textrm{i}} }{ \sum_{\textrm{i}} \abs{\vec{p}_{\textrm{i}}}^{2}  }~\mbox{.}
\end{equation}
The sphericity of the event is defined in terms of the two largest eigenvalues of this tensor, $\lambda_2$ and $\lambda_3$:
\begin{equation}
	S=\frac{3}{2}({\lambda_2 + \lambda_3}).
\end{equation}
Similar to that of transverse thrust, the transverse sphericity is defined in terms
of the transverse components only: 
\begin{equation}
	S_{\textrm{xy}} = \sum_{i} \begin{bmatrix}
  p_{\textrm{x,i}}^2 & p_{\textrm{x,i}}p_{\textrm{y,i}} \\
  p_{\textrm{x,i}}p_{\textrm{y,i}} & p_{\textrm{y,i}}^2 \\
 \end{bmatrix}
\end{equation}
and
\begin{equation}
	S_{\textrm{T}} = \frac{2\lambda^{xy}_{2}}{\lambda^{xy}_{1} + \lambda^{xy}_{2}}~\mbox{,}
\end{equation}
where again $\lambda^{xy}_{2}>\lambda^{xy}_{1}$ are the two eigenvalues of
$S_{\textrm{xy}}$.

Sphericity is essentially a measure of the summed $p_\textrm{T}^2$ with respect to the
event axis. 
Sphericity lies between $0<S_T<1$, where a 2-jet event corresponds to $S_\textrm{T}= 0$ and an
isotropic event to $S_\textrm{T}=1$.

\section{Analysis setup}
\label{sec:setup}

$Z$-bosons are reconstructed from oppositely charged muon pairs with invariant mass 
in the range from $66$ to $116$ GeV.
The muons are required to have \pT$ > 25$~GeV and $|\eta| < 2.5$.
Charged particles with \pT$>0.5$~GeV and $|\eta| < 2.5$ are used to construct underlying event
and event shape observables, and the muons from the $Z$-boson decay are excluded. Jets are reconstructed 
using the anti-\kt algorithm~\cite{Cacciari:2008gp}
with a radius parameter of $R=0.4$, and are required to have  \pT $> 25$~GeV and $|\eta| < 4.5$. 
A center-of-mass energy of $13$~TeV is assumed.

\begin{figure}[h!]
  \centering
  \includegraphics[width=0.5\textwidth]{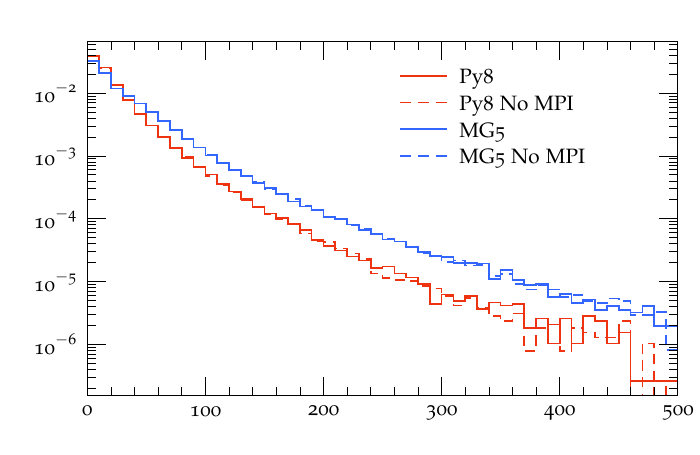}\qquad
  \caption{$Z$-boson \pT distribution with MPI on and off
  for LO and multileg setups.}
  \label{fig:zpt}
\end{figure}

The \pythiaeight~\cite{Sjostrand:2007gs} generator (v8.226) was 
used to generate events at leading order (LO), while 
the \mg~\cite{Alwall:2014hca}
generator (v260) was used to generate events $Z$-boson events with up to three (QCD) extra jets (multileg). 
Events generated with \mg were
showered with \pythiaeight using the MLM~\cite{Hoche:2006ph} matching scheme. In both cases,
the Monash tune~\cite{Skands:2014pea} 
with NNPDF2.3LO parton density function~\cite{Ball:2012cx} was used. 
Each of these samples were generated again with turning MPI off.
For events  generated with \mg, the additional 
jets corresponding to ISR originate from hard scatter, as opposed to from parton shower in \pythiaeight with the tune used.
So the former are expected to be more energetic, as evidenced indirectly from $Z$-boson \pT distribution
shown in \FigRef{fig:zpt}.
One million events was generated in each case. No detector simulation or smearing
is applied.
The Rivet analysis toolkit~\cite{Buckley:2010ar} has been used.

\section{Results}
\label{sec:res}

\begin{figure}[h!]
  \centering
  \includegraphics[width=0.4\textwidth]{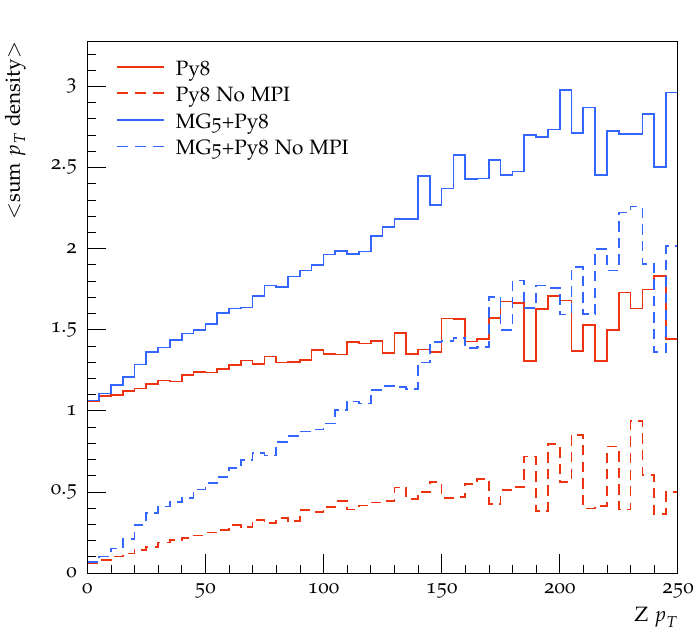}\qquad
   \includegraphics[width=0.4\textwidth]{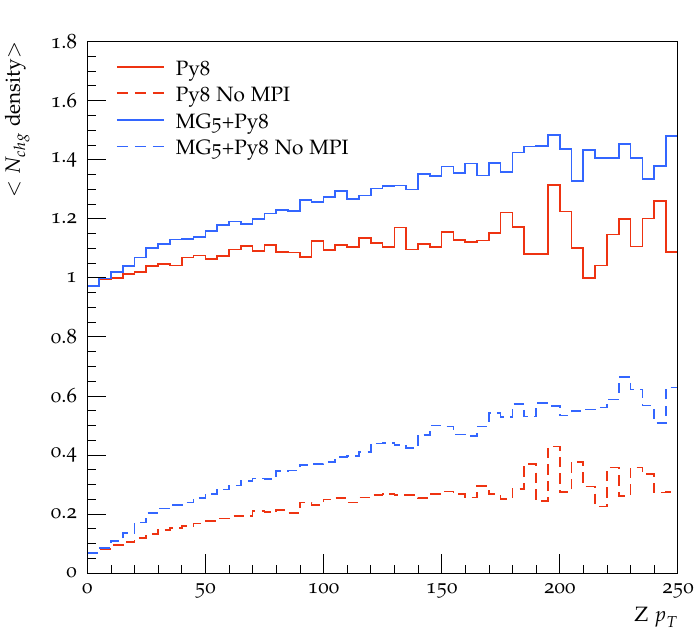} \\
  \caption{Charged particle scalar sum \pT density (left) and multiplicity density (right) as functions of Z \pT in toward region, both with MPI on and off
  for LO and multileg setups.}
  \label{fig:ue1}
\end{figure}

The toward region is one of the most important from the point of view of the UE in $Z$-boson events. 
Comparing the charged particle
sum \pT and multiplicity densities as functions of $Z$-boson \pT obtained from \pythiaeight
and \mg+\pythiaeight generators show that ISR hard jets indeed lead to much higher charged particle activity, as can be seen in \FigRef{fig:ue1}. 
The effect of
turning off MPI is roughly a constant decrease in the activity. So the target is to find events where the difference
due to ISR jets is negligible.

\begin{figure}[h!]
  \centering
  \includegraphics[width=0.4\textwidth]{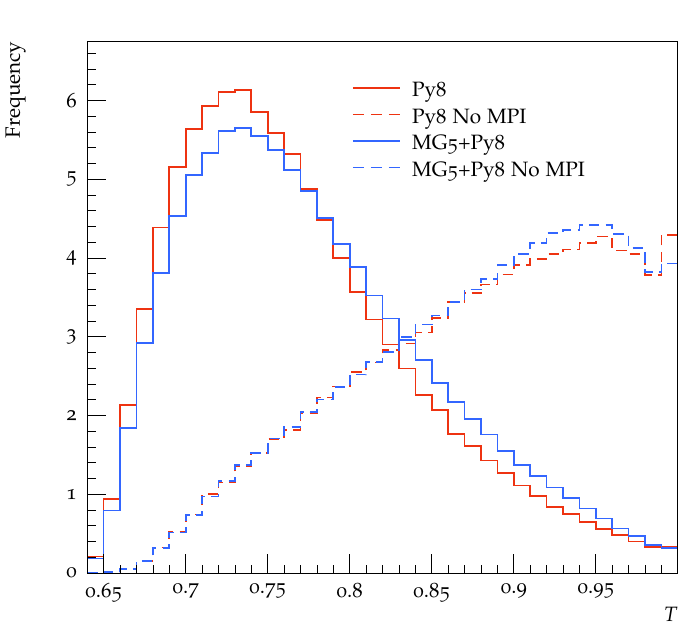}\qquad
   \includegraphics[width=0.4\textwidth]{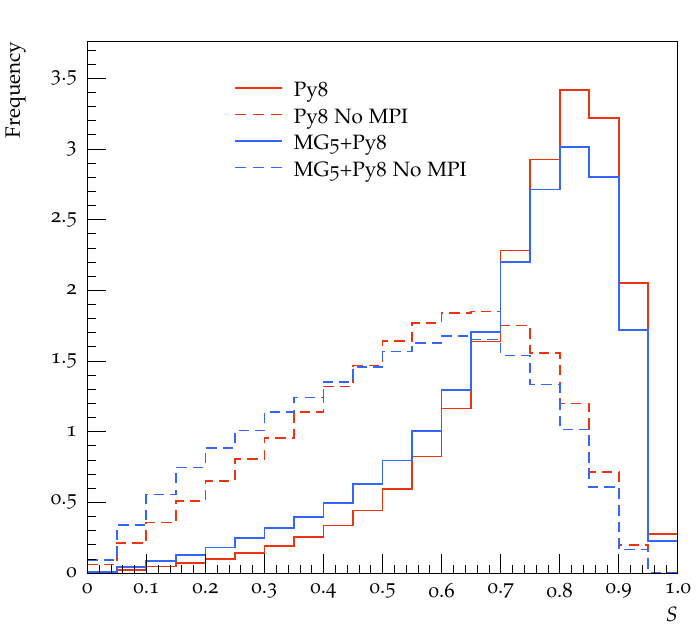} \\
  \caption{Transverse thrust (left) and transverse sphericity (right) distributions  both with MPI on and off
  for LO and multileg setups.}
  \label{fig:es1}
\end{figure}

Looking at \FigRef{fig:es1}, depicting the inclusive distribution of transverse thrust and sphericity, it is clear
that MPI results in making events more spherical.

\begin{figure}[h!]
  \centering
  \includegraphics[width=0.4\textwidth]{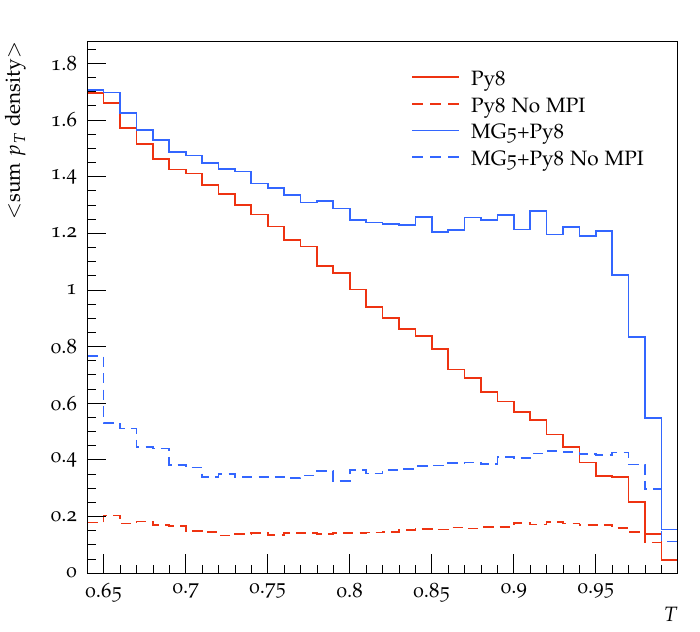}\qquad
   \includegraphics[width=0.4\textwidth]{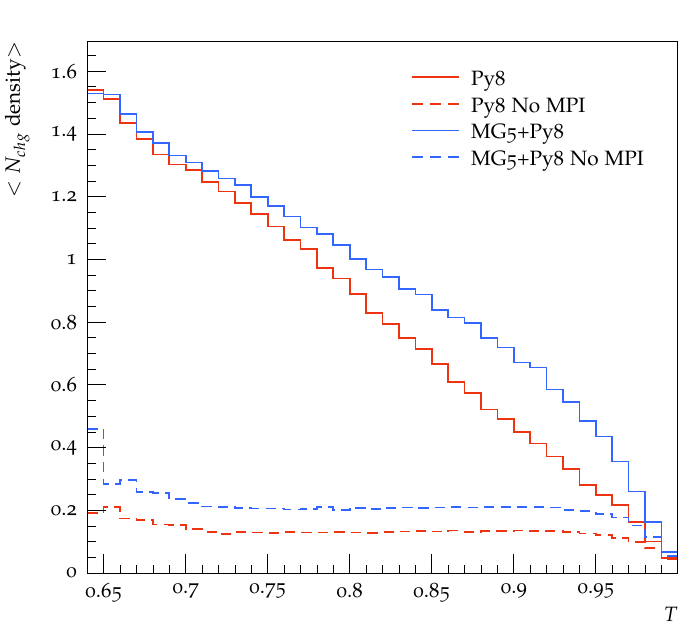} \\
  \caption{Charged particle scalar sum \pT density (left) and multiplicity density (right) as functions of 
  transverse thrust in the toward region, both with MPI on and off
  for LO and multileg setups.}
  \label{fig:thr}
\end{figure}

Now combining the two sets of observables, \FigRef{fig:thr} shows 
the charged particle scalar sum \pT and multiplicity densities as a function of transverse thrust. A number of
interesting observations can be made from these plots:

\begin{itemize}

\item The charged particle multiplicity or scalar sum \pT (density) 
activity predicted by \mg+\pythiaeight, 
slowly decreases with increase of transverse thrust values.
This indicates that the spherical events have more activity. 
In fact, the activity is almost the 
same for \mg+\pythiaeight and \pythiaeight for most isotropic events.
That must mean that events with more activity are more spherical, while events with less activity are more dijet-like.
For \mg this is true for multiplicity, but not true for sum \pT, which can be
inferred by a more steeply falling distribution to higher values of thrust (dijet-like) for scalar
sum $p_T$ distribution compared to charged particle multiplicity.
So that implies that in the toward region \mg gives
more soft particles compared to \pythiaeight, causing events to be more spherical, but the spherical events 
have lower sum \pT.

\item The transition to more dijet-like events is sharper for \pythiaeight, resulting from the absence of 
ISR jets. In other words, fewer ISR jets in \pythiaeight causes much more isotropic events.

\item Without MPI,  both the distributions are mostly flat over the all transverse thrust range, 
showing that the effect of ISR jets is independent of
topology. The activity without MPI is non zero for \pythiaeight, indicating presence of softer ISR jets.

\item The charged particle multiplicity distribution is noticeably different from the sum \pT distribution.
The activity falls for \mg +\pythiaeight even with the presence of jets from hard scatter, indicating MPI results in
softer particles than those jets.

\end{itemize}

These features were seen to come almost exclusively from events with more than one jet.
Transverse region profiles show the same trend as toward, but are more affected by ISR jets.

\begin{figure}[h!]
  \centering
  \includegraphics[width=0.4\textwidth]{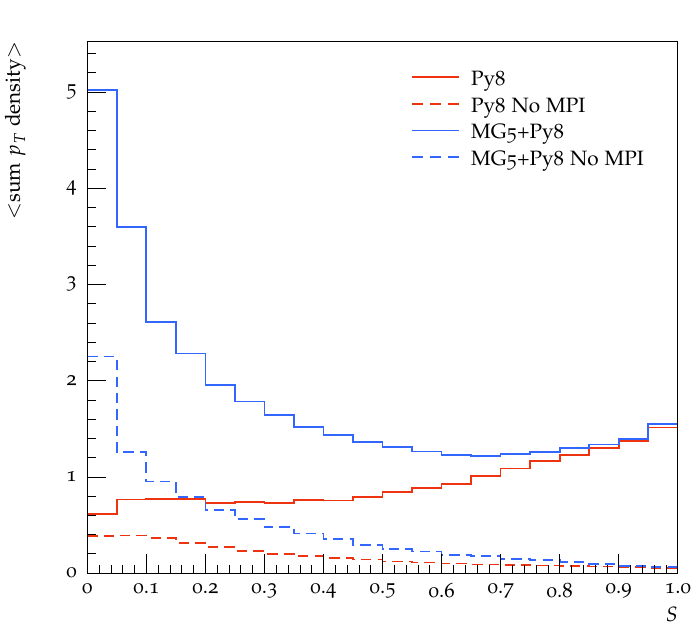}\qquad
   \includegraphics[width=0.4\textwidth]{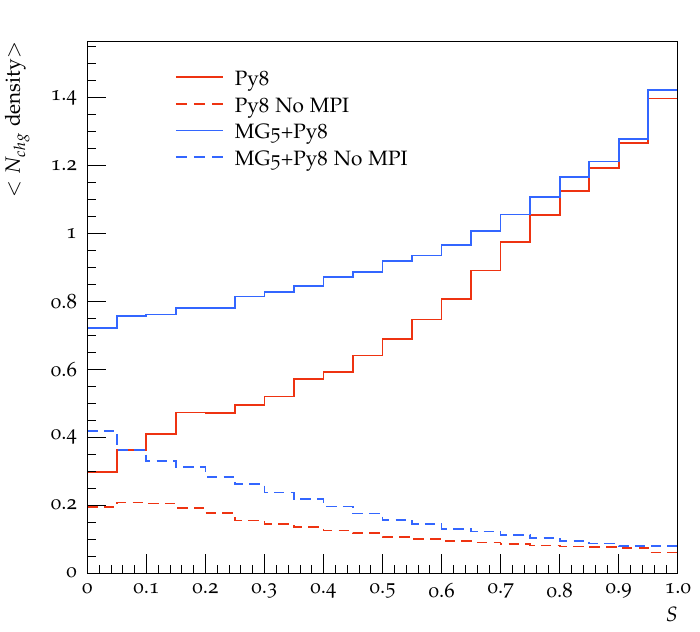} \\
  \caption{Charged particle scalar sum \pT density (left) and multiplicity density (right) as functions of transverse sphericity in toward region, both with MPI on and off for LO and multileg setups.}
  \label{fig:sph}
\end{figure}

In \FigRef{fig:sph}, the same distributions for the UE observables as a function of transverse sphericity 
are constructed.
The features are qualitatively similar,
but slightly less prominent, indicating that the transverse thrust is more sensitive to the effect we are probing.

\FloatBarrier

\section{Conclusions}
\label{sec:smm}

From these results, it can be inferred that a measurement of the UE observables for transverse thrust $<0.75$, or transverse
sphericity $>0.65$ would select a class of events where the effect of ISR jets is smaller. These can then help
to constrain MPI in a much cleaner way in $Z$-boson events. 
This approach will have a few advantages over performing
the measurement in jet multiplicity bins. 
The jets usually have a \pT threshold of $25$ GeV, while constructing 
event shape observables using charged particles, we use more information from the events.
Additionally event shape observables also represent
geometrical directions of activity in the event,
 which a pure jet counting would not do as well. Finally, 
 not using jets will not make us susceptible to additional systematic uncertainties in an 
 experimental measurement, associated with jet energy scale and resolution, for example.

\section{Acknowledgements}
\label{sec:ack}

D.K would like to thank the African Institute of Mathematical Sciences (AIMS), Cape Town, for 
letting D.S.R work on his structured masters research project.
D.K is supported by National Research Foundation, South Africa in forms of CSUR and Incentive funding.


\bibliographystyle{ws-ijmpa}
\bibliography{ref}

\begin{thebibliography}{10}
\expandafter\ifx\csname urlstyle\endcsname\relax
  \providecommand{\doi}[1]{doi:\discretionary{}{}{}#1}\else
  \providecommand{\doi}{doi:\discretionary{}{}{}\begingroup
  \urlstyle{rm}\Url}\fi

\bibitem{Aaltonen:2010rm}
 CDF Collaboration (T.~Aaltonen {\em et~al.}), {\em Phys. Rev. D} {\bf 82},
  034001  (2010), \href{http://arxiv.org/abs/1003.3146}{{\ttfamily
  arXiv:1003.3146 [hep-ex]}}, \doi{10.1103/PhysRevD.82.034001}.

\bibitem{Aad:2014jgf}
 ATLAS Collaboration (G.~Aad {\em et~al.}), {\em Eur. Phys. J. C} {\bf 74},
  3195  (2014), \href{http://arxiv.org/abs/1409.3433}{{\ttfamily
  arXiv:1409.3433 [hep-ex]}}, \doi{10.1140/epjc/s10052-014-3195-6}.

\bibitem{Chatrchyan:2012tb}
 CMS Collaboration (S.~Chatrchyan {\em et~al.}), {\em {Eur. Phys. J. C}} {\bf
  {72}},   2080  ({2012}), \href{http://arxiv.org/abs/1204.1411}{{\ttfamily
  arXiv:1204.1411 [hep-ex]}}, \doi{10.1140/epjc/s10052-012-2080-4}.

\bibitem{Sirunyan:2017vio}
 CMS Collaboration (A.~M. Sirunyan {\em et~al.}), {\em JHEP} {\bf 07},   032
  (2018), \href{http://arxiv.org/abs/1711.04299}{{\ttfamily arXiv:1711.04299
  [hep-ex]}}, \doi{10.1007/JHEP07(2018)032}.

\bibitem{Field:2002vt}
CDF Collaboration, R.~D. Field, { The underlying event in hard scattering
  processes}, in {\em Proceedings, APS / DPF / DPB Summer Study on the Future
  of Particle Physics (Snowmass 2001), Snowmass, Colorado, 30 Jun - 21 Jul
  2001\/},  {\em eConf} {\bf C010630} (2001), p. 501.
\newblock \href{http://arxiv.org/abs/hep-ph/0201192}{{\ttfamily
  arXiv:hep-ph/0201192 [hep-ph]}}.

\bibitem{PhysRevD.38.3419}
G.~Marchesini and B.~R. Webber, {\em Phys. Rev. D} {\bf 38},   3419  (1988),
  \doi{10.1103/PhysRevD.38.3419}.

\bibitem{PhysRevD.57.5787}
J.~Pumplin, {\em Phys. Rev. D} {\bf 57},   5787  (1998),
  \doi{10.1103/PhysRevD.57.5787}.

\bibitem{Berger:2010xi}
C.~F. Berger, C.~Marcantonini, I.~W. Stewart, F.~J. Tackmann and W.~J.
  Waalewijn, {\em JHEP} {\bf 04},   092  (2011),
  \href{http://arxiv.org/abs/1012.4480}{{\ttfamily arXiv:1012.4480 [hep-ph]}},
  \doi{10.1007/JHEP04(2011)092}.

\bibitem{Banfi:2010xy}
A.~Banfi, G.~P. Salam and G.~Zanderighi, {\em JHEP} {\bf 06},   038  (2010),
  \href{http://arxiv.org/abs/1001.4082}{{\ttfamily arXiv:1001.4082 [hep-ph]}},
  \doi{10.1007/JHEP06(2010)038}.

\bibitem{Aad:2016ria}
 ATLAS Collaboration (G.~Aad {\em et~al.}), {\em Eur. Phys. J. C} {\bf 76},
  375  (2016), \href{http://arxiv.org/abs/1602.08980}{{\ttfamily
  arXiv:1602.08980 [hep-ex]}}, \doi{10.1140/epjc/s10052-016-4176-8}.

\bibitem{Chatrchyan:2013tna}
 CMS Collaboration (S.~Chatrchyan {\em et~al.}), {\em Phys. Lett. B} {\bf 722},
  238  (2013), \href{http://arxiv.org/abs/1301.1646}{{\ttfamily arXiv:1301.1646
  [hep-ex]}}, \doi{10.1016/j.physletb.2013.04.025}.

\bibitem{Cuautle:2014yda}
E.~Cuautle, R.~Jimenez, I.~Maldonado, A.~Ortiz, G.~Paic and E.~Perez  (2014),
  \href{http://arxiv.org/abs/1404.2372}{{\ttfamily arXiv:1404.2372 [hep-ph]}}.

\bibitem{Cacciari:2008gp}
M.~Cacciari, G.~P. Salam and G.~Soyez, {\em JHEP} {\bf 04},   063  (2008),
  \href{http://arxiv.org/abs/0802.1189}{{\ttfamily arXiv:0802.1189 [hep-ph]}},
  \doi{10.1088/1126-6708/2008/04/063}.

\bibitem{Sjostrand:2007gs}
T.~Sj{\"o}strand, S.~Mrenna and P.~Skands, {\em Comput. Phys. Commun.} {\bf
  178}, 852  (2008), \href{http://arxiv.org/abs/0710.3820}{{\ttfamily
  arXiv:0710.3820 [hep-ph]}}, \doi{10.1016/j.cpc.2008.01.036}.

\bibitem{Alwall:2014hca}
J.~Alwall, R.~Frederix, S.~Frixione, V.~Hirschi, F.~Maltoni, O.~Mattelaer,
  H.~S. Shao, T.~Stelzer, P.~Torrielli and M.~Zaro, {\em JHEP} {\bf 07},   079
  (2014), \href{http://arxiv.org/abs/1405.0301}{{\ttfamily arXiv:1405.0301
  [hep-ph]}}, \doi{10.1007/JHEP07(2014)079}.

\bibitem{Hoche:2006ph}
S.~Hoeche, F.~Krauss, N.~Lavesson, L.~Lonnblad, M.~Mangano, A.~Schalicke and
  S.~Schumann, { Matching parton showers and matrix elements}, in {\em HERA and
  the LHC: A Workshop on the implications of HERA for LHC physics: Proceedings
  Part A\/},  (2005), pp. 288--289.
\newblock \href{http://arxiv.org/abs/hep-ph/0602031}{{\ttfamily
  arXiv:hep-ph/0602031 [hep-ph]}}.

\bibitem{Skands:2014pea}
P.~Skands, S.~Carrazza and J.~Rojo, {\em Eur. Phys. J. C} {\bf 74},   3024
  (2014), \href{http://arxiv.org/abs/1404.5630}{{\ttfamily arXiv:1404.5630
  [hep-ph]}}, \doi{10.1140/epjc/s10052-014-3024-y}.

\bibitem{Ball:2012cx}
R.~D. Ball {\em et~al.}, {\em Nucl. Phys. B} {\bf 867}, 244  (2013),
  \href{http://arxiv.org/abs/1207.1303}{{\ttfamily arXiv:1207.1303 [hep-ph]}},
  \doi{10.1016/j.nuclphysb.2012.10.003}.

\bibitem{Buckley:2010ar}
A.~Buckley, J.~Butterworth, L.~Lonnblad, D.~Grellscheid, H.~Hoeth, J.~Monk,
  H.~Schulz and F.~Siegert, {\em Comput. Phys. Commun.} {\bf 184}, 2803
  (2013), \href{http://arxiv.org/abs/1003.0694}{{\ttfamily arXiv:1003.0694
  [hep-ph]}}, \doi{10.1016/j.cpc.2013.05.021}.

\end{thebibliography}

\end{document}